\documentclass[twocolumn,showpacs,preprintnumbers,amsmath,amssymb]{revtex4}
\usepackage{tabularx,graphicx}

\usepackage{color}
\usepackage{hyperref}
\hypersetup{
    colorlinks=true,
    linkcolor=blue,
    filecolor=blue,      
    urlcolor=blue,
}

\usepackage{color}

\usepackage{ulem}   % to strike things out
\begin{document}
%\documentstyle[aps]{revtex}
%\documentstyle[preprint,aps]{revtex}
%\begin{document}

\newcommand{\beq}{\begin{equation}}
\newcommand{\eeq}{\end{equation}}
\newcommand{\beqn}{\begin{eqnarray}}
\newcommand{\eeqn}{\end{eqnarray}}
\newcommand{\bmath}{\begin{subequations}}
\newcommand{\emath}{\end{subequations}}
\newcommand{\bra}[1]{\langle #1|}
\newcommand{\ket}[1]{|#1\rangle}

%\draft
\title{Meissner effect in nonstandard superconductors}

\author{J. E. Hirsch$^{a}$  and F. Marsiglio$^{b}$ }
\address{$^{a}$Department of Physics, University of California, San Diego,
La Jolla, CA 92093-0319\\
$^{b}$Department of Physics, University of Alberta, Edmonton,
Alberta, Canada T6G 2E1}

\begin{abstract} 
It was recently pointed out that so-called ``superhydrides'', hydrogen-rich materials that appear to become superconducting at high temperatures and pressures, 
exhibit physical properties that are different from both 
conventional and unconventional standard type I and type II superconductors \cite{hm,dc}.
Here we consider magnetic field expulsion in the first material in this class discovered in 2015, sulfur hydride \cite{eremetssh}.
A nuclear resonant scattering experiment has been interpreted as demonstration that the Meissner 
effect takes place in this material \cite{nrs,nrs2}. Here we point out that the observed effect, under the assumption that the system is in
thermodynamic equilibrium, implies a
Meissner pressure \cite{londonbook} in this material that is {\it much larger} than that of standard  
superconductors. This suggests that hydride superconductors are qualitatively different from the
known standard superconductors   {\it if} they are superconductors.

\end{abstract}
\maketitle 
\section{introduction}
The 2015 discovery of high temperature superconductivity in pressurized sulfur hydride \cite{eremetssh} 
($H_3S$)
with critical temperature up to 203 K was the first
  of several metal hydrides recently reported to be superconducting at high temperatures and pressures
  between 100 GPa and 250 GPa.
These include phosphorous hydride at above 100 K \cite{eremetsp}, 
lanthanum hydride at 250 K \cite{eremetslah}, above 260 K \cite{hemleylah} and above
550 K \cite{hemleylah2}, yttrium hydride at 243 K \cite{yttrium2,yttrium,yttriumdias}, thorium hydride at 161 K \cite{thorium},
 lanthanum-yttrium ternary hydrides at 253 K \cite{layh10}, carbonaceous sulfur hydride at room 
 temperature \cite{roomt} and cerium hydride above 110 K \cite{ceh}.
 For all these materials, resistance versus temperature curves exhibit sharp drops at temperatures that have been interpreted
 as superconducting transition temperatures, and application of large magnetic fields shifts these transition
 temperatures
 to lower values. These materials have been characterized as strongly type II superconductors, with
 upper critical fields in the range 60-150~T \cite{review2}.
 
 However, in recent work we \cite{hm} and others \cite{dc} have  argued   that features of the resistive transition in a magnetic field
 appear to be in conflict with the behavior seen in standard superconductors,
 explained by the conventional theory of superconductivity \cite{tinkham}, and hence that these materials,
 if they are superconductors, are `nonstandard superconductors' \cite{hm}.

To establish that superconductivity exists in these materials it would be important to show that they
expel magnetic fields, i.e. the Meissner effect; alternatively even if somewhat less compelling, that they do not
allow penetration of magnetic fields, i.e. magnetic field exclusion. In Refs. \cite{nrs,nrs2}, it has been claimed that 
magnetic field exclusion in 
sulfur hydride has been detected through a novel nuclear resonance scattering experiment.
However, here we show that a standard superconductor, whether conventional or
unconventional, would not show this behavior in thermodynamic equilibrium. We conclude that, under this assumption,  this observation provides further evidence that 
sulfur hydride, and by inference other
hydride superconductors, are either nonstandard superconductors \cite{hm}, or they are not superconductors.
In a separate paper \cite{fluxtrap} we consider the possibility that the measurements reported in ref. \cite{nrs} did not reflect
the state of the superconductor in thermodynamic equilibrium.

\section{the NRS experiment}
In Ref. \cite{nrs}, a $2.6-\mu m$ thick foil of tin was immersed into the $H_3S$ specimen \cite{h3s}.
The tin was enriched to $95\%$ with the $^{119}Sn$ isotope to serve as the sensor.
The sensor monitors the magnetic field via the magnetic interaction at the $^{119}Sn$ nucleus as detected by
nuclear resonant scattering (NRS) of synchrotron radiation. The presence of magnetic field at tin nuclei was
identified by quantum beats in the time spectra of NRS \cite{nrs3}.

As a control, the measurements were conducted simultaneously with two diamond anvil cells (DACs). 
One contained the $H_3S$ sample, the other contained only $H_2$. Both had identical $^{119}Sn$-enriched
foils, and were in the same applied magnetic field and temperature. It was found that the spectra measured
were similar for both samples above 100 K, but differed markedly for temperatures below 50 K.
While the control sample response was consistent with having a magnetic field in its interior 
equal to the applied magnetic field, $H_{ext}\sim0.68$ T, the response of the $H_3S$ sample indicated
that the magnetic field in the interior of the sample dropped to zero for temperatures 50 K and below
when the magnetic field was applied perpendicular to the sample, and to about a third of the
applied field when the magnetic field was applied parallel to the sample.

The experiment did not attempt to detect the Meissner effect, i.e. magnetic field expulsion, as the paper states, but rather 
magnetic field exclusion \cite{expulsion}. The sample was first cooled in the absence of a magnetic field and
subsequently the magnetic field was applied.
The authors concluded that the experiment showed that the magnetic field was excluded from the interior
of the sample at low temperatures due to the Meissner effect, and therefore the sample had to be
in the superconducting state at low temperatures.

In the following sections we show that a standard superconductor in thermodynamic equilibrium would not have excluded the magnetic field
in the conditions of the experiment.  

        \begin{figure} []
 \resizebox{8.5cm}{!}{\includegraphics[width=6cm]{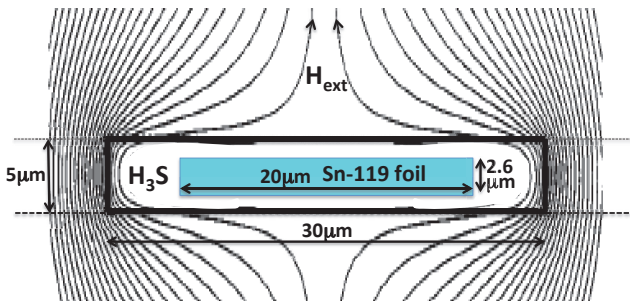}} 
 \caption {Geometry of the experiment with field perpendicular to the sensor film \cite{nrs}.
  The $Sn$ foil is assumed to be 20 $\mu m$ in diameter and 2.6 $\mu m$ in height, 
 the sulfur hydride superconducting sample is assumed to be 30 $\mu m$ in diameter and 5 $\mu m$ in  height   \cite{nrs}.
}
 \label{figure2}
 \end{figure}

\section {magnetic field exclusion}

Let us first consider the geometry where the applied magnetic field is perpendicular to the $Sn$ foil, because
it is in this geometry where the largest deviation from the expected behavior of standard superconductors occurs.

Figure~1 shows the geometry, also shown in Figure S6 of Ref.~\cite{nrs}. The external magnetic field had magnitude
$H_{ext}=0.68$ T. The figure shows the assumed exclusion of the magnetic field from the
sample interior. It is immediately obvious that the magnetic field at the edges of the sample is much larger
than the applied field, due to demagnetization. 
For a cylindrical geometry with aspect ratio $height/radius=0.333$, as shown in Fig.~1, 
the demagnetizing factor is approximately $N_z \approx 0.727$ \cite{demag}. This indicates that the magnetic field
at the edges of the sample is a factor $1/(1-N_z)=3.7$ times larger than the applied field.
So the magnetic field at the edge of the sample is 
\beq
H_{edge}=2.5 \ {\rm T} .
\eeq

If this is a type I superconductor, it would imply that the thermodynamic critical field $H_c$ is larger than $2.5$~T.
This  would be unprecedented. The largest known critical fields for standard type I superconductors are of order
0.05 T, i.e. 50 times smaller. If this is a type II superconductor in thermodynamic equilibrium,  
it would imply that the lower critical field $H_{c1}$ is larger than $2.5$~T. This would be even more unprecedented,
since $H_{c1}<H_c$ within the standard theory of superconductivity.

In Ref.~\cite{eremetssh}, the upper critical field $H_{c2}$ was estimated to be between $60$ and $80$~T. 
Let us assume $H_{c2}=68$~T, corresponding to a zero temperature coherence length $\xi=2.20\ nm$. 
From magnetization measurements, it was concluded in Ref.~\cite{eremetssh} that the lower critical field is approximately 
$H_{c1}\sim0.03$~T. 
The applied field is then 1/100 of $H_{c2}$ and much larger than
$H_{c1}$. 
From this,  the London penetration depth was estimated to be $\lambda_L=125\ nm$ \cite{eremetssh}. 
Alternatively, Talantsev estimated a value $\lambda_L=189\ nm$ from self-field critical current data \cite{tsal}.
We will use the value of $\lambda_L$ of Ref.~\cite{eremetssh} for definiteness; note that our conclusions would not change using
a larger $\lambda_L$ value.

It is clear that if this was a standard superconductor that will reach its thermodynamic equilibrium state, the applied field which is much larger than $H_{c1}$ would penetrate throughout the superconductor, no matter what the geometry, in particular in the two configurations used in \cite{nrs}, magnetic field perpendicular to the $Sn$ foil (Fig.1) and parallel to it.
For a sample of cross-sectional area $A$ perpendicular to the magnetic field, the number of vortices $N_v$ is determined by the equation
\beq
H_{ext}A=N_v\phi_0 .
\eeq
The upper critical field is given by
\beq
H_{c2}=\frac{\phi_0}{2\pi \xi^2}
\eeq
so the area per vortex is
\beq
a_v=\frac{A}{N_v}=    2\pi\xi^2\frac{H_{c2}}{H_{ext}}
\eeq
so for $H_{ext}/H_{c2}=0.01$,
\beq
a_v=\pi (\sqrt{200}\xi)^2
\eeq
so the distance between vortex cores is approximately
\beq
d_v=2\sqrt{200}\xi=62\ nm,
\eeq
which is half the estimated London penetration depth. This indicates that the magnetic field is {\it nearly uniform} in the mixed state of this superconductor,
with magnitude in the range
\beq
e^{-d_v/(2\lambda_L)}H_{ext}<H<H_{ext}
\eeq
or 
\beq
0.78H_{ext}<H<H_{ext} .
\eeq

So  using the parameters for this superconductor inferred by the authors of Ref.~\cite{eremetssh} and assuming standard superconductivity,
we conclude that in both geometries used in Ref.~\cite{nrs}, applied magnetic field perpendicular and parallel to the $Sn$ film,
the magnetic field would be uniform or nearly uniform inside the sample and inside the $Sn$ film  when the sample is in the superconducting state
in thermodynamic equilibrium.
However this is inconsistent with the experimental results presented in Fig.~4 of Ref.~\cite{nrs}, that indicate that the magnetic field drops
to zero in one geometry and to one third of the applied value in the other geometry, for temperatures below $50$~K, 
as is shown  in Fig.~2.

        \begin{figure} []
 \resizebox{8.5cm}{!}{\includegraphics[width=6cm]{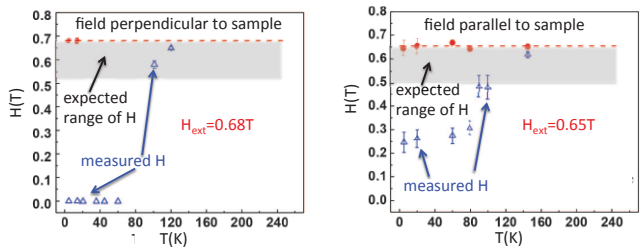}} 
 \caption {Comparison of experimental results from the NRS experiment of Ref.~\cite{nrs} with 
 what would be expected if sulfur hydride is a standard superconductor that reaches thermodynamic equilibrium, for applied
 magnetic field perpendicular and parallel  to the $Sn$ foil.
The blue triangles
 and red dots  show
 the values of the magnetic field at the $Sn$ foil inferred from the experimental data 
 for the $H_3S$ sample and the control $H_2$ sample respectively in Ref.~\cite{nrs}.
  The grey strips indicate the range of magnetic fields where the blue triangles should be located if
  $H_3S$ was a standard superconductor according to Eq.~(8).
}
 \label{figure2}
 \end{figure} 

But let us assume that for some unknown  reason the magnetic field is indeed zero in the interior of the sample,
as claimed by the authors of ref. \cite{nrs}.
What would be the current necessary so that the field does not penetrate? A lower bound to that current
would be a current circulating near the surface that generates a magnetic field $H_{edge}=0.68$~T at the center
of the sample. For a circular loop of current of radius $a$ with current $I$ circulating, the magnetic field at the center is
\beq
B=\frac{2\pi I}{ca}
\eeq
which for $B=0.68$~T and $a=15 \ \mu m$ as shown in Fig.~1 yields $I=16$~Amp. Assuming the current circulates within the
London penetration depth $\lambda_L=125\ nm$ of the surface and sample thickness $5\ \mu m$ as shown in
Fig.~1 yields for the current density
\beq
J=2.6\times 10^{9}\frac{Amp}{cm^2} .
\eeq
Therefore, the critical current density would have to be larger than the value Eq. (10). 
Critical current densities for all known standard superconductors are about 2 orders of magnitude lower than
that value. This then  suggests that if $H_3S$ is a superconductor, it is a nonstandard superconductor,
as defined in Ref.~\cite{hm}. Note that we also found in Ref.~\cite{hm}, based on analysis of the resistive
transition in a magnetic field, that the critical current densities of nonstandard superconductors are much higher
than those of standard superconductors.

\section{nonstandard superconductivity in $H_3S$}
To further explore  the possibility that if $H_3S$ is a superconductor it is a nonstandard superconductor, 
let us ignore the values of $\lambda_L$ reported in Refs.~\cite{eremetssh,tsal} mentioned above and instead infer its value using
the standard theory of superconductivity.
As discussed in the previous section, the total magnetic field at the edge of the sample would be
$2.5$~T if the field doesn't penetrate. Within the standard theory of superconductivity this implies that the lower critical field $H_{c1}$ is larger
than $2.5$~T even at temperatures around
$100$~K where the magnetic field starts to be excluded. Let us explore the implications of
this.
For simplicity we will assume $H_{c1}=2.5$~T at $T=0$, although in reality it would have to be
even larger if it is $2.5$~T at $\sim 100$~K.

Using the expression for the lower critical field \cite{tinkham,approx}
\beq
H_{c1}(0)=\frac{\phi_0}{4\pi\lambda_L(0)^2}ln\left({\lambda_L(0) \over \xi(0)}\right)
\eeq
we have for the London penetration depth at zero temperature
\beq
\lambda_L(0)= \xi(0) \sqrt{\frac{H_{c2}(0)}{2H_{c1}(0)} ln\left({\lambda_L(0)\over \xi(0)}\right)} 
\eeq
and with $H_{c2}(0)=68$~T, $H_{c1}(0)=2.5$~T 
\beq
\frac{\lambda_L(0)}{\xi(0)}=3.69 \sqrt{ln(\frac{\lambda_L(0)}{\xi(0)}}
\eeq
and we obtain $\lambda_L(0)=4.54 \ \xi(0)=10.0\ nm$. 
So the material would have to be a 
 weakly type II superconductor, with $\kappa=\lambda_L(0)/   \xi(0)=4.5$, to exclude the applied
magnetic field as found in the NRS experiment, instead of 
$\kappa >50$ as inferred in Refs.~\cite{eremetssh,tsal}.  
In facy, it was argued in Ref. \cite{roomt} that a similar material in this class, CSH, may be such a weakly type II superconductor, with $\kappa$
as low as $1.7$ \cite{diasreply}.

However, this would be contrary to all expectations for this material (and for CSH \cite{hm}) based on the standard theory of
superconductivity \cite{eremetssh,review2, tsal}. In addition, it 
would imply that the thermodynamic critical field, given by 
\beq
H_c(T)=\frac{\phi_0}{2\sqrt{2}\pi\lambda_L(T)\xi(T)}
\eeq
has the value 
\beq
H_c(0)=10.6\ {\rm T}
\eeq
which is more than an order of magnitude larger than is found for any standard superconductor, and implies an
enormous condensation energy.  This in turn exerts an enormous `Meissner pressure' 
$H_c^2/(8\pi)$ \cite{londonbook} which is necessary to account for the experimental results seen in 
Fig.~2 in the configuration of Fig.~1.

Within the standard theory of superconductivity the thermodynamic critical field obeys the
relation
\beq
\frac{H_c^2(0)}{8\pi}= \frac{1}{2}g(\epsilon_F)\Delta^2(0)
\eeq
where $\Delta(0)$ is the energy gap at zero temperature and $g(\epsilon_F)$ is the density of states per spin
at the Fermi energy. So we would have to assume that this relation
which holds for standard superconductors fails for the superhydrides for unknown reasons.
Alternatively, this value of $H_c(0)$ would imply that either the density of states or the energy gap is much larger than in standard
superconductors. Assuming the standard BCS relation between energy gap and critical temperature,
 the density of states that would be implied by the value Eq.~(16) for the critical 
field is
\beq
g(\epsilon_F)=\frac{0.586 \ states}{spin-eV\AA^3} .
\eeq
This is an enormous density of states. For comparison, using the standard theory the density of states
of sulfur hydride was estimated to be $0.019$ $states/(spin-eV /\AA^3)$ \cite{review2},
30 times smaller.

Assuming  the density of states is given by the free electron 
expression with effective mass $m^*$  we have
\beq
g(\epsilon_F)=\frac{0.0206}{spin-eV\AA^2}\frac{m^*}{m_e}n^{1/3}
\eeq
with $n$ the number density.  For the density of states given by Eq. (17) this yields for the Wigner Seitz radius
\beq
r_s=0.022\frac{m^*}{m_e}\AA .
\eeq
The $H-H$ distance in sulfur hydride at pressures above 150 GPa is approximately $1.4 \ \AA$.
Assuming $r_s=0.7\ \AA$ yields
\beq
\frac{m^*}{m} \approx 32.
\eeq
However, the effective mass enhancement resulting from the electron-phonon interaction is expected to be
only around a factor of 3. The theoretical calculations that claim to explain the observed values of
$T_c$ in $H_3S$  \cite{th0,th1,th2,th3,th4,th5,th6} are not compatible with an effective mass enhancement as given 
by Eq.~(20).

Note also that if the London penetration depth is $\lambda_L\sim 10.0 \ nm$, as given by this analysis rather than
$125\ nm$, the critical current would have to be larger than $3.2\times 10^{10} Amp/cm^2$, 
even more anomalous than the value given by Eq.~(10).

We conclude that if the
experimental results reported in Ref.~\cite{nrs} reflect the properties of $H_3S$ in thermodynamic equilibrium, $H_3S$ cannot be explained by the standard theory of superconductivity. In that case, $H_3S$   would be  a novel nonstandard superconductor that excludes magnetic fields
that are  two orders of magnitude larger than what the standard theory predicts it can exclude. 
Presumably this is would be  true for all the other nonstandard superconductors \cite{hm}, i.e. all other
hydrogen-rich materials that superconduct at high temperatures under high pressures.

\section{discussion}

If sulfur hydride in thermodynamic equilibrium truly excludes magnetic fields of the magnitude claimed in the geometries used in this
experiment, it would imply that its superconductivity is very nonstandard.  It would have a ``Meissner effect on steroids''.
This new property of nonstandard superconductors 
would be consistent with what was found in Ref.~\cite{hm}, that 
nonstandard superconductors should have critical currents that are orders of magnitude higher than
those of standard superconductors. A new theory of the electrodynamics of these materials would be  required to describe their
very unusual  properties.
These properties would lend them very useful for practical applications such as levitating trains.

However, if these materials were so different from standard superconductors we also have to ask:  why would it be
that standard BCS-Eliashberg theory appears to be able to accurately predict  their  transition temperatures as claimed in the
literature  \cite{th0,th1,th2,th3,th4,th5,th6,th7,th8,th9,th10,th11,th12}?
We suggest that `superflexibility' \cite{rainer} may have something to do with it, whether or not these
materials are superconductors.

There is also the possibility  that the experiment reported in Ref.~\cite{nrs} is flawed for
some reason. If so, it would not provide   supporting evidence to the claim that sulfur hydride under pressure
 is a high temperature superconductor \cite{eremetssh,review2}, contrary to what is generally assumed \cite{googlessh3}. This would
 cast further doubt on the claim that superhydrides are high temperature superconductors,
adding to the arguments given in \cite{hm,dc,eu}.

We will consider the alternative possibility that the measurements in \cite{nrs} do not reflect the properties of $H_3S$ in thermodynamic equilibrium,
and its implications, in a separate paper \cite{fluxtrap}.

 \begin{acknowledgments}
JEH is grateful to M. L. Cohen for stimulating discussions that motivated his interest in this study. We are grateful to D. Semenok for discussions and
for sharing results of not yet published experiments. 
FM 
was supported in part by the Natural Sciences and Engineering
Research Council of Canada (NSERC) and by an MIF from the Province of Alberta.  

\end{acknowledgments}


\begin{references}
\bibitem{expulsion} The abstract of the paper mistakenly claims that the results demonstrate
that the magnetic field {\it ``is expelled from the volume''} of the sensor.
\bibitem{h3s} At the time of the experiment, it was believed that superconducting hydrogen sulfide 
          was $H_2S$; presently it is generally agreed it is $H_3S$ \cite{review2} that has the
          higher transition temperature.
\bibitem{approx} This is an approximate relation, but will suffice for our purposes.
\end{references}
 \end{document}